\documentclass{pasj00}

\usepackage{xspace,graphicx}
\newcommand{\NH}{\mbox{$N_{\rm H}$}}        
\newcommand{\suzaku}{{\sl Suzaku}\xspace}
\newcommand{\swift}{{\sl Swift}\xspace}
\newcommand{\cd}{{$10^{22}$\,cm$^{-2}$}\xspace}
\newcommand\feHe{{Fe {\sc xxv}}\xspace}
\newcommand\feH{{Fe {\sc xxvi}}\xspace}
\def\ss17{{\sl SS73~17}\xspace}

\begin{document}

\SetRunningHead{Smith et al.}{The Symbiotic System SS73 17 Seen with Suzaku}
\Received{2007/06/29}
\Accepted{2001/01/01}

\title{The Symbiotic System SS73 17 Seen with Suzaku} 

\author{%
  Randall K. \textsc{Smith}\altaffilmark{1,2}
  Richard \textsc{Mushotzky}\altaffilmark{1}
  Koji \textsc{Mukai}\altaffilmark{3,4}
  Tim \textsc{Kallman}\altaffilmark{1}
  Craig B. \textsc{Markwardt}\altaffilmark{5,6}
  and
  Jack \textsc{Tueller}\altaffilmark{1}}
\altaffiltext{1}{NASA/Goddard Space Flight Center, Greenbelt, MD 20771}
\email{rsmith@milkyway.gsfc.nasa.gov}
\altaffiltext{2}{Department of Physics and Astronomy, The Johns Hopkins
  University, Baltimore, MD 21218}
\altaffiltext{3}{CRESST and X-ray Astrophysics Laboratory, NASA/GSFC,
  Greenbelt, MD 20771} 
\altaffiltext{4}{Department of Physics, University of Maryland,
  Baltimore County, 1000 Hilltop Circle, Baltimore, MD 21250}
\altaffiltext{5}{CRESST and Astroparticle Physics Laboratory, NASA/GSFC,
  Greenbelt, MD 20771} 
\altaffiltext{6}{Department of Astronomy, University of Maryland,
  College Park, MD 20742} 


\maketitle

\begin{abstract}
We observed with Suzaku the symbiotic star SS73 17, motivated by the
discovery by the INTEGRAL satellite and the Swift BAT survey that it
emits hard X-rays.  Our observations showed a highly-absorbed X-ray
spectrum with $N_{\rm H} > 10^{23}$\,cm$^{-2}$, equivalent to $A_{\rm
V} > 26$, although the source has B magnitude 11.3 and is also bright
in UV.  The source also shows strong, narrow iron lines including
fluorescent Fe K as well as \feHe and \feH.  The X-ray
spectrum can be fit with a thermal model including an absorption
component that partially covers the source.  Most of the equivalent
width of the iron fluorescent line in this model can be explained as a
combination of reprocessing in a dense absorber plus reflection off a
white dwarf surface, but it is likely that the continuum is partially
seen in reflection as well.  Unlike other symbiotic systems that show
hard X-ray emission (CH Cyg, RT Cru, T CrB, GX1+4), SS73 17 is not
known to have shown nova-like optical variability, X-ray flashes, or
pulsations, and has always shown faint soft X-ray emission.  As a
result, although it is likely a white dwarf, the nature of the compact
object in SS73 17 is still uncertain.  SS73 17 is probably an extreme
example of the recently discovered and relatively small class of hard
X-ray emitting symbiotic systems.
\end{abstract}

\section{Introduction}      

\ss17 has long been known as a moderately bright (V$\sim 10$)
symbiotic star --- until recently, undistinguished from many other
such systems \citep{SS73}(SS73).  ROSAT observations from 1992
showed the source was a faint ($F_X(0.5-2$\,keV$) =
3.6\times10^{-13}$\,erg cm$^{-2}$s$^{-1}$) soft X-ray source
\citep{Bickert96}, but beyond that little was known at UV or higher
energies.  In 2005, however, this system was independently discovered
to be a hard X-ray source by both INTEGRAL (IGRJ10109-5746;
\citet{Revnivtsev06a}) and Swift (Swift J101103.3-574814;
\citet{Atel669}).  Before these sources were identified as being
associated either with each other or with \ss17, we successfully
proposed for \suzaku observations of the Swift source, with the goal
of finding potential members of the newly-identified class of
``highly-absorbed X-ray binaries'' discovered by INTEGRAL
\citep{Kuulkers05}.  Our choice of Swift J101103.3-574814 was driven
by its hard X-ray flux, as seen by the Swift BAT ($F_X(20-60{\rm keV})
= 1.6\times10^{-11}$\,ergs cm$^{-2}$s$^{-1}$), and its low Galactic
latitude ($l,b = 282.9^{\circ},-1.3^{\circ}$).

Although \citet{Atel669} noted that this source is only $30'$\ from
GRO J1008-57, a transient XRB with a pulse period of 93.6 s, ROSAT
PSPC observations which also detected the pulsations confirmed GRO
J1008-57's position to within $15''$ \citep{IAU5877}.  Based on the
Swift XRT and ROSAT PSPC data, we can be confident that SWIFT
J1010.1-5747, IGR J10109-5746 and \ss17 are all names for a single
symbiotic system that is distinct from GRO J1008-57.

A symbiotic star consists of a red giant and a hot blue companion,
frequently a white dwarf accreting material from the red giant wind
\citep{Kenyon86}.  Although many symbiotic stars emit soft X-rays
\citep{Murset97}, only small number of those with white dwarf
companions are known to emit hard X-rays, including CH Cyg
\citep{Ezuka98}, RT Cru \citep{Atel715}, and T CrB \citep{Atel669}.
In addition to hard X-rays, each of these have also shown optical
flares or novae at some point
\citep{Deutsch74,Atel528,CordovaMason84}.  Another type of symbiotics,
the ``symbiotic low mass X-ray binaries,'' with M giant/neutron star
systems such as GX1+4 and 4U1954+31 \citep{Mattana06} also shows hard
X-ray emission, frequently with a periodicity of minutes to hours.
\ss17 is particularly intriguing because it could {\sl only}\ be
identified as unusual in hard X-rays, since its soft X-ray emission is
nondescript.  Its optical emission history is patchy, appearing in
catalogs roughly every few decades since its initial listing as CD-57
3057.  If in fact it has little optical variability, \ss17 may be
first example of a new class of symbiotic/X-ray binary that has a
significant hard X-ray flux {\bf without}\ significant soft X-ray
emission or optical outbursts.

\section{Previous Observations}

\ss17 was listed as an ``interesting'' star in the SS73 catalog, where
it was described as a M3ep + OB binary system.  \citet{Henize76} noted
the star shows Balmer lines in emission through H 10.  Subsequently,
\citet{Pereira03} obtained spectra from a number of SS73 stars, and
identified \ss17 as a Mira-type system, including a normal-type M4
giant star, with emission lines of H$\alpha$\ and H$\beta$\ along with
TiO absorption bands.  They suggested that Henize's observations were
likely done at object maximum phase, since they did not see such
strong emission lines.  Although not observed by IUE, the source
saturated the Swift UVOT in all 4 Swift observations, with a UVW1
magnitude greater than 11.3 (based on the quick-look
data\footnote{http://heasarc.gsfc.nasa.gov/docs/swift/sdc/data\_products.html}). 

The optical survey of variable stars by \citet{Pojmanski03} found weak
evidence for a 577 day period for \ss17, with V magnitude variations
between 9.68-9.89.  Unlike other hard X-ray symbiotic systems (CH Cyg,
RT Cru, T CrB), \ss17 has not shown any sign of a optical brightening
or ``slow nova'' in its sparse historical record.  The distance to
\ss17 is not known, although it was observed by the Hipparcos
satellite as part of the Tycho catalog.  The measured parallax was
$94.9\pm30.2$\,mas ($10.5^{+4.9}_{-2.5}$\,pc), but this value is not
credible for an M4III star with $V\sim9.7$.  \citet{Allen00}(p.406)
notes that the absolute V magnitude of Mira variables is $\approx
0.0040 P - 2.6$.  For P=577 days, this corresponds to $M_V = -0.3$, or
a maximum distance (assuming no absorption) of 1 kpc.  The line of
sight absorption to \ss17 is also unknown, but assuming the M giant
dominates emission in the J and K bands, $E(J-K)=0.56$\ and $A_V
\approx 3.3$ \citep{Allen00}(p.158), equivalent to N$_{\rm H} \approx
6\times10^{21}$\,cm$^{-2}$, implying a distance of $\sim 220$\,pc.
Finally, \ss17 is $\sim 1.5$\,mag fainter than the apparently similar
CH Cyg system (at $D = 245\pm50$\,pc, \citet{Perryman97}), suggesting
a distance of 500 pc.  Combining these results, we estimate \ss17's
distance to be in the range $250-1000$\,pc.

As noted above, \ss17 was observed both by the ROSAT All-sky survey
and in a 7 ksec pointed observation, finding a weak source in both
cases.  In addition, EXOSAT observed a $45^{'}\times45^{'}$\ field
containing \ss17 with the (non-imaging) Medium Energy (ME) detector
(sequence \#1340), although the data quality was noted as ``poor.''  
We downloaded the standard spectrum and response from the HEASARC
archive and fit it with XSPEC to an absorbed bremsstrahlung spectrum
as the data quality did not justify a more complex model.  The
best-fit model had a flux of $F_X(2-10 {\rm keV}) \sim 10^{-11}$\,erg
cm$^{-2}$s$^{-1}$\ with a temperature greater than 3 keV and \NH $\sim
1-5\times10^{22}$\,cm$^{-2}$.  The EXOSAT observation also showed that
no source in the field did not show any strong variability during the
5 hour observation.  Although we cannot be certain the EXOSAT spectrum is
from \ss17, the spectrum and flux is consistent with that conclusion.
The other other known nearby X-ray source, GRO J1008-57, is transient
and, when on, was significantly brighter than the EXOSAT source.

\section{Observations and Analysis}

\ss17 was observed by \suzaku for 20 ksec on June 5, 2006 (obsid
01055010).  The pointing direction was chosen to center the source on
the HXD detector, which has the effect of reducing the effective area
of the XIS detectors by 10\% due to vignetting.  All four XIS
detectors were in standard imaging mode.  The data were processed
using version 1.2 of the standard \suzaku pipeline software.

We extracted all events within $4.34'$\ of \ss17 for each of the four
XIS detectors to create the source spectra.  The lightcurve of both
the XIS and HXD events showed no sign of flares either in the complete
event list or the source events, so all events were kept.  We
generated response matrices for the XIS detectors using version
2007-05-14 of {\tt xisrmfgen}, and used the standard effective area
files for HXD-nominal pointings using a 6 mm ($4.34'$) extraction
circle\footnote{Available at
ftp://legacy.gsfc.nasa.gov/caldb/data/suzaku/xis/\ cpf/ae\_xi\{0,1,2,3\}\_hxdnom6\_20060615.arf}.
The XIS background data was taken from a circular region with no apparent
sources that was offset from both the source and the corner
calibration sources.  

The source was not bright enough to appear in the HXD GSO, so we only
consider the HXD PIN detector in this paper.  Since the observation
was done after the W0 PIN diode voltage was reduced to 400V from 500V
but before this mode was calibrated, we eliminated these events from
our dataset and used only the W123 events\footnote{See
http://heasarc.gsfc.nasa.gov/docs/suzaku/analysis/pinbias.html}.  We
then used the PIN response matrix appropriate for the W123 diodes,
correcting for dead time using the {\tt hxddtcor}\ routine.  We used
the PIN background events file (v1.2) generated by the Suzaku HXD
team\footnote{Available at
http://www.astro.isas.jaxa.jp/suzaku/analysis/hxd/\ pinnxb/pinnxb\_ver1.2\_w123/401055010/ae401055010hxd\_pinnxb\_cl.evt.gz}
to make the PIN non-X-ray background (NXB) spectrum.  The cosmic X-ray
background was included in the spectral modeling itself, at the level
found by \citet{Boldt87}.

\section{Results}

\begin{figure}
\begin{center}
\FigureFile(80mm,80mm){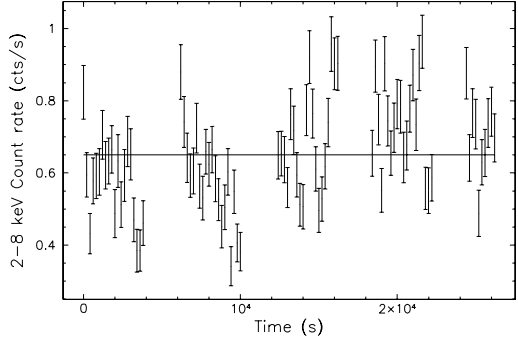}
\end{center}
\caption{\suzaku XIS (all 4 coadded) lightcurve between 2-8 keV; the
  average rate is shown as a straight line.  Although some
  short term variability is seen, there are no obvious periodicities.
\label{fig:ltcrv}}
\end{figure}

\begin{figure}
\begin{center}
\FigureFile(80mm,80mm){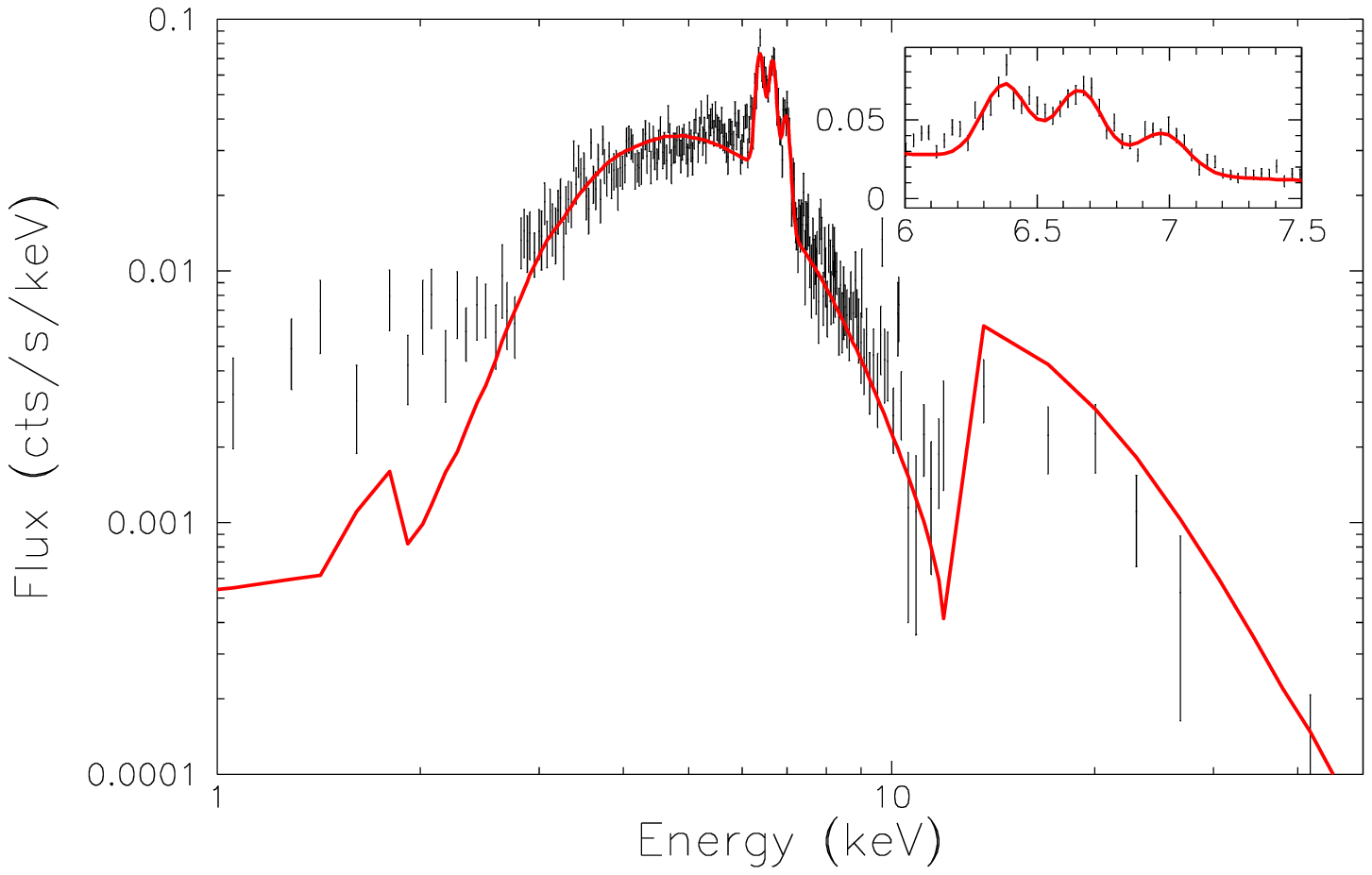}
\FigureFile(80mm,80mm){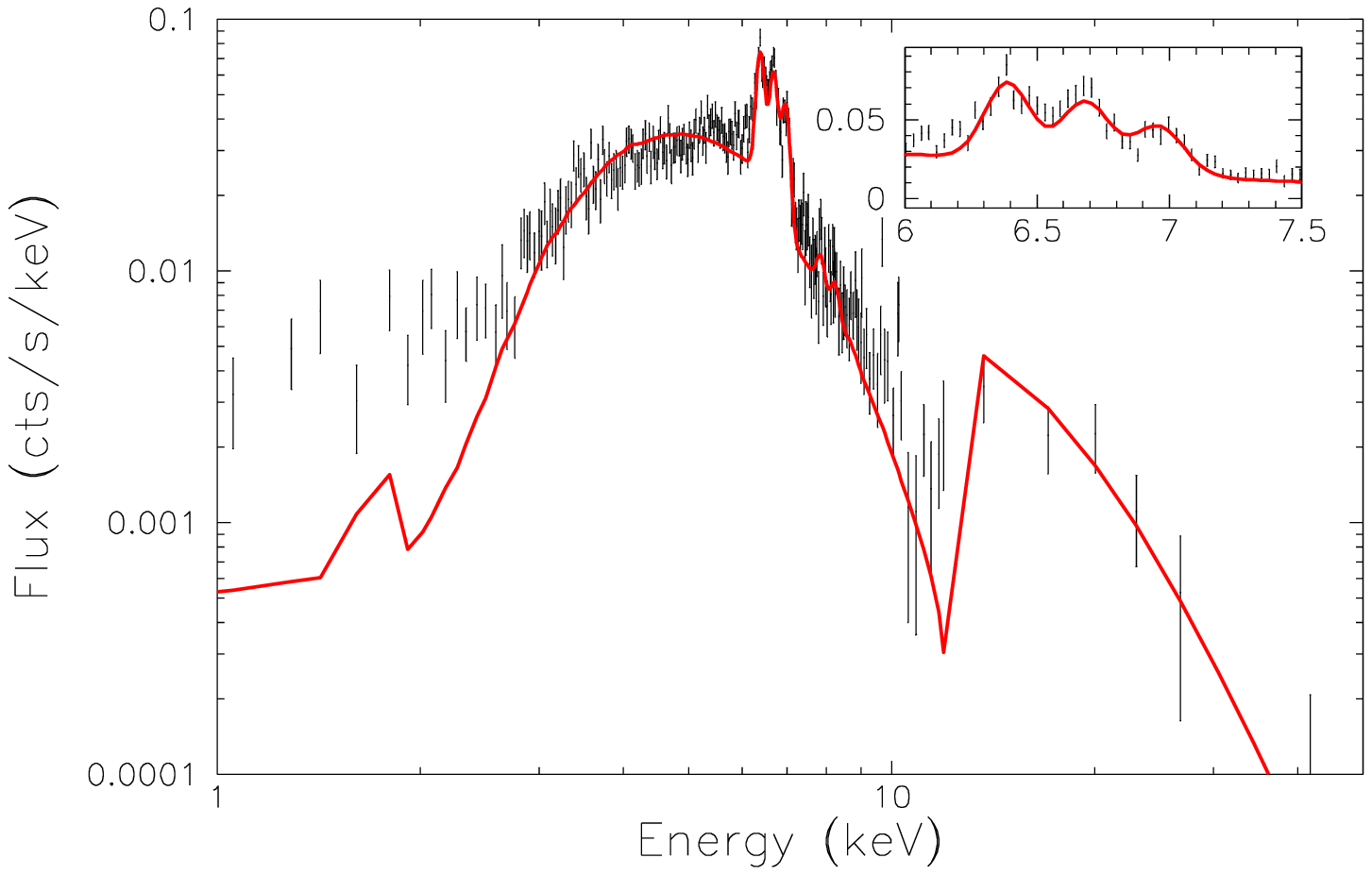}
\end{center}
\caption{[Top] \suzaku XIS (all 4 coadded) and HXD PIN
  background-subtracted spectra between 1-50 keV, with best-fit
  absorbed power law plus three emission lines model.  The inset
  spectrum shows the 6-8 keV range where the iron lines appear.
  [Bottom] Same, for an absorbed thermal model plus a single emission
  line to reproduce the 6.4 keV Fe fluorescence line.  In both cases,
  the soft emission is not reproduced by the model.\label{fig:spec}}
\end{figure}

Figure~\ref{fig:spec} shows \ss17's X-ray spectrum as observed by the
\suzaku XIS and HXD and demonstrates its primary features: 
\begin{enumerate}
\item weak emission below 2 keV
\item a strongly absorbed hard X-ray spectrum
\item three narrow, strong emission lines at 6.38, 6.66, and 6.98 keV.
\end{enumerate}
\suzaku's energy calibration between 6-8 keV is accurate to $\pm6$\,eV
\citep{Koyama07}, so the three emission lines can be identified as an
iron K fluorescence line as well as \feHe and \feH emission.  However,
the exact position and strength of the lines depends upon the shape of
the underlying continuum.  In addition to the spectral analysis, we
show \ss17's lightcurve (between 2-8 keV) in Figure~\ref{fig:ltcrv}.
The background-subtracted count rate averaged 0.7 cts/s, fluctuating
between 0.5-1 cts/s without an apparent period.  This stability is
similar to what was seen in the earlier EXOSAT ME observation, and
suggests that the process generating the X-ray emission (likely
accretion) is relatively steady.

\begin{figure}
\begin{center}
\FigureFile(80mm,80mm){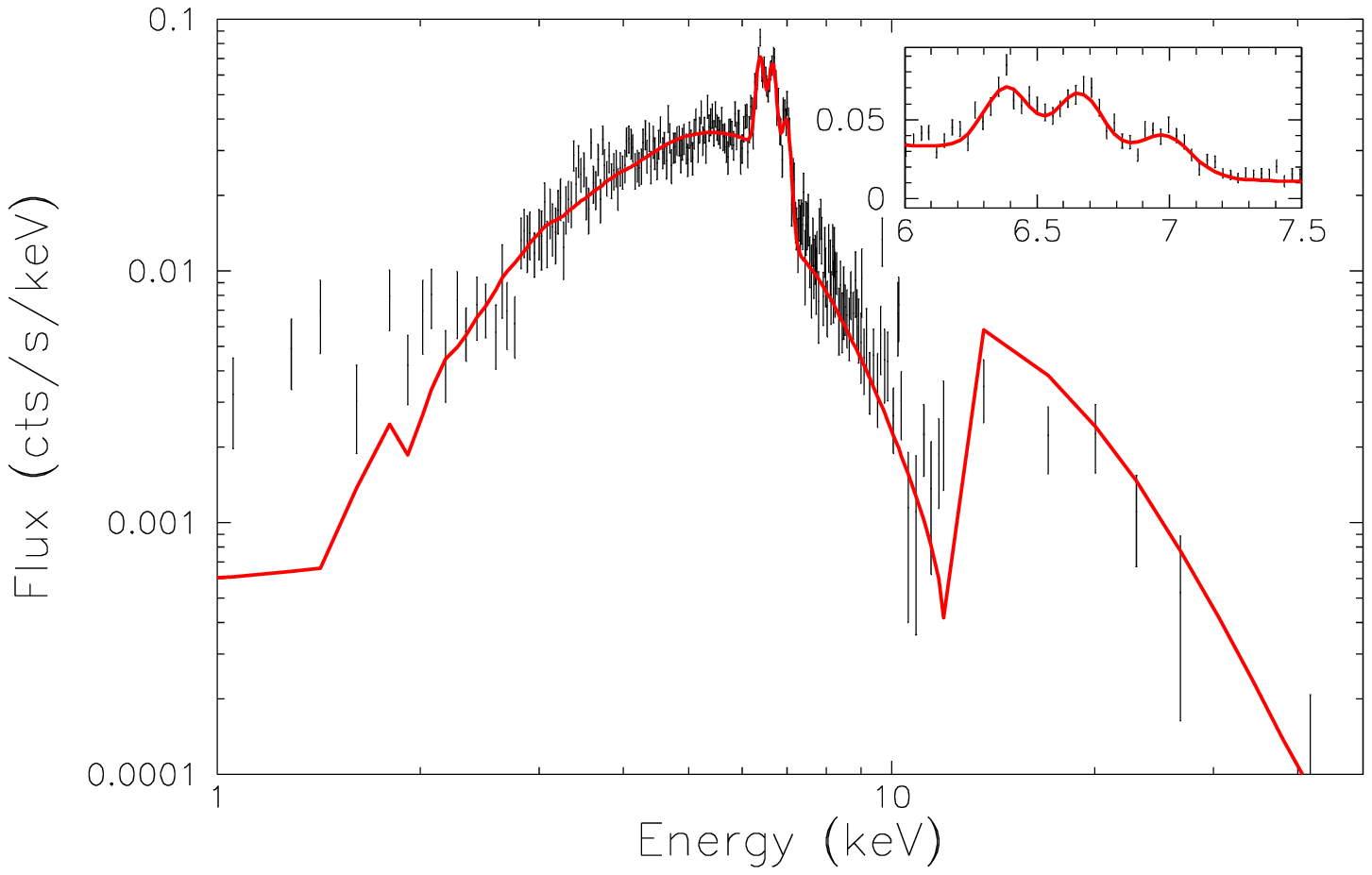}
\FigureFile(80mm,80mm){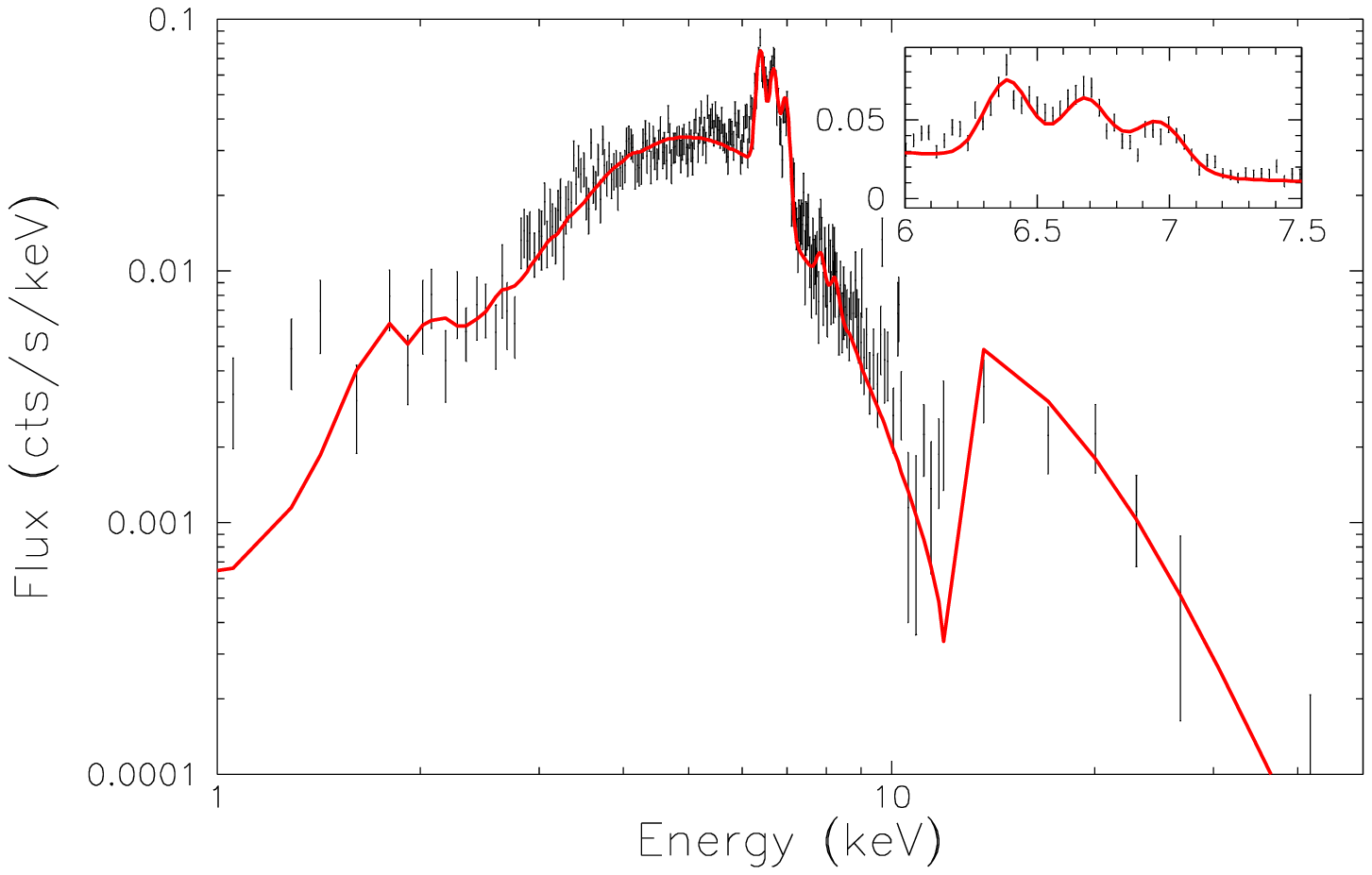}
\end{center}
\caption{[Top] \suzaku XIS (all 4 coadded) and HXD PIN
  background-subtracted spectra between 1-50 keV, with best-fit
  model that includes both a normal and a partial-covering absorber
  model combined with a power law and three emission lines model.  The
  inset spectrum shows the 6-8 keV range where the iron lines appear.
  The soft emission is not well-modeled.  [Bottom] Same, for an
  similarly-absorbed thermal model with a single additional emission
  line to reproduce the 6.4 keV Fe fluorescence line.  
  \label{fig:absspec}}
\end{figure}

We considered four different models for the X-ray spectrum, based on
both a synchrotron (power law) origin and a thermal model for the
continuum emission.  The synchrotron model assumes that the emission
comes from relativistic electrons near the compact object, perhaps
emitted as a yet-unseen jet.  In contrast, the thermal model supposes
that the emission is due primarily to a hot stream of accreting
material falling onto the compact object.  In both cases we added a
Gaussian to model the iron fluorescence line, which presumably arises
from a combination of fluorescence in the absorbing material and
scattering from the surface of the hot star or an accretion disk.
In the power law model, we also added two additional Gaussians to
model the \feHe and \feH emission features.  In the thermal model we
assumed these latter two lines arose from the plasma itself.  All four
models included at least one absorption component; we also considered
models with a second absorber that only partially covered (noted with
``PC'' in Table~\ref{tab:fits}) the source.  This second absorber is
motivated by the low interstellar absorption ($A_V \approx 3.3, N_H
\approx 6\times10^{21}$\,cm$^{-2}$) to the Mira variable companion,
which strongly suggests the X-ray absorption is extremely local to the
X-ray source in the binary system.  The soft X-ray emission between
1-2 keV is best fit by the partial-covering thermal model, although
even this model underpredicts the emission in this band.  However,
when the 0.5-2 keV band is fit with a simple thermal model absorbed by
Galactic emission, the total flux is only $F_X(0.5-2) =
(2.3_{-0.6}^{+0.9})\times10^{-14}$\,erg cm$^{-2}$s$^{-1}$.  This soft
X-ray flux is an order of magnitude fainter than either the ROSAT
all-sky or pointed observation result, although these are rather
uncertain values as only $\sim 200$\,total counts were observed
\citep{Bickert96}.  The soft X-ray variation suggests the partial
absorber periodically changes to allow more photons to escape the
system, evidence that the local absorber does not enshroud the X-ray
source.  Our results for all four models are shown in
Figures~\ref{fig:spec} and \ref{fig:absspec}\ and in
Tables~\ref{tab:fits} and \ref{tab:FeK} for the model parameters and
iron line strengths, respectively.  All quoted errors show 90\%
confidence intervals.  Assuming a distance in the range
$0.25-1.0$\,kpc, the $F_X(2-10{\rm keV}) \approx 8.5\times10^{-12}$\
absorbed flux implies a luminosity of $L_X(2-10) =
(0.6-10)\times10^{32}$\,erg/s.  The unabsorbed flux depends upon the
model used; in the ``Abs PC TH'' case, it is
$(1.5-23)\times10^{32}$\,erg/s.

\begin{longtable}{llllllll}
\caption{Spectral Fit Results}\label{tab:fits}
\hline \hline
Model    &  \NH       &Covering&\NH     &kT          &$\Gamma$
&$\chi^2_{\nu}$&$F_X(2-10$\,keV)\\ 
         & \cd        &Fraction&\cd     &keV         &
&&$10^{-12}$\,erg cm$^{-3}$s$^{-1}$\\ \endfirsthead \hline \endhead
\endhead
  \hline
\endfoot
  \hline
\endlastfoot
  \hline
Abs PL   &$15.8\pm1.1$&\dots   &\dots   &\dots       &$1.78\pm0.04$&1.3&$8.4_{-1.1}^{+0.4}$\\
Abs TH   &$16.3\pm0.7$&\dots   &\dots   &$9.3\pm0.5$&\dots         &1.3&$8.5_{-1.9}^{+0.1}$\\
Abs PC PL&$10.7\pm2.0$&$85\pm2$&$41_{-7}^{+10}$&\dots&$2.6\pm0.1$  &1.2&$9.1_{-3.6}^{+0.1}$\\
Abs PC TH&$ 3.5\pm1.4$&$92_{-5}^{+3}$&$18\pm1.5$&$9.4\pm0.5$&\dots &1.2&$8.5\pm0.2$\\ \hline
\end{longtable}

The high energy ($E= 10-50$\,keV) result from the PIN is at the limit
of detectability, but is in line with other results.  INTEGRAL
detected with source with significance $\sigma=5.57$\ and $F_X(17-60)
= 1.3\pm0.3$mCrab \citep{Revnivtsev06a}.  We find (using the partial
covering thermal fit) $F_X(17-60) = (1.16\pm0.04)\times10^{-11}$\,ergs
cm$^{-2}$s$^{-1} \approx 1$\,mCrab.  The Swift BAT detection of the
source found $F_X(20-60) = 1.6\times10^{-11}$\,ergs cm$^{-2}$s$^{-1}
\approx 1.4$\,mCrab, in line with both results.  Any differences
between our results and those of INTEGRAL and Swift are well within
the systematic uncertainties in the PIN background subtraction (which
are at the $\sim5\%$\ of background level, or $\approx 0.6$\,mCrab).

\begin{longtable}{lllllll}
\caption{Emission Lines from \ss17}\label{tab:FeK}
\hline \hline
Model    &\multicolumn{2}{c}{Fe K}
         &\multicolumn{2}{c}{\feHe}
	 &\multicolumn{2}{c}{\feH}\\
         & Flux ($10^{-5}$ & Eq. Width 
         & Flux ($10^{-5}$ & Eq. Width 
         & Flux ($10^{-5}$ & Eq. Width \\
         & ph cm$^{-2}$s$^{-1}$) & keV 
         & ph cm$^{-2}$s$^{-1}$) & keV 
         & ph cm$^{-2}$s$^{-1}$) & keV \\ \hline \endfirsthead \hline \endhead
\endhead
  \hline
\endfoot
  \hline
\endlastfoot
  \hline
Abs PL   & $6.8\pm0.6$ &$0.25\pm0.05$& $7.1\pm0.6$ &$0.23\pm0.05$& $3.7\pm0.5$&$0.17\pm0.05$\\
Abs TH   & $7.1_{-0.5}^{+0.7}$ &$0.27\pm0.05$&  \dots      &  \dots      & \dots  & \dots \\
Abs PC PL& $9.5\pm1.5$&$0.19_{-0.03}^{+0.06}$&$9.2\pm1.3$&$0.19_{-0.04}^{+0.06}$&$4.1\pm0.8$&$0.12_{-0.05}^{+0.07}$\\
Abs PC TH& $7.7_{-0.8}^{+0.05}$&$0.26_{-0.04}^{+0.06}$&  \dots      &  \dots      &\dots  & \dots \\\hline
\end{longtable}

\section{Discussion and Conclusions}

\citet{Murset97} classified 16 symbiotics seen with ROSAT into three
types: (1) supersoft emission from the atmosphere of the hot star, (2)
emission from an optically-thin plasma with $kT\sim0.2$\,keV, or (3)
an ill-defined catch-all of relatively hard X-ray sources.  They found
only two sources in class (3), the X-ray binary GX1+4 and Hen 3-1591
(which they suspected might also contain a neutron star).  Although
not in the M\"urset et al. (1997) survey, the ROSAT observation of
\ss17 suggests it would likely have been placed in category (2),
albeit with some uncertainty due to the large column density (N$_{\rm
H} = 1.8\times10^{22}$\,cm$^{-2}$) required to fit the ROSAT data.

However, subsequent X-ray observatories have discovered hard X-rays
from the symbiotic systems CH Cyg, RT Cru, T Crb in addition to \ss17
\citep{Kennea07} whose emission does not fit with the physical
mechanisms described by \citet{Murset97}.  Curiously, the X-ray
spectrum of \ss17 above 2 keV appears be quite similar to the Polar
magnetic Cataclysmic Variable (mCV) star AM Her.  \citet{Ezuka99} fit
the 1993 ASCA observation of AM Her using an Abs PC bremsstrahlung
model, finding $kT = 10.0$\,keV with $58\pm10$\% of the source
absorbed by a column density of $(36\pm7)\times10^{22}$\,cm$^{-2}$,
$20\pm3$\% absorbed with a column density of
$(3.6\pm0.6)\times10^{22}$\,cm$^{-2}$, with the remaining area
effectively unabsorbed.  The fluorescent iron line equivalent width in
this model was $0.17\pm0.04$\,keV, and the bolometric luminosity of
the system was $L_X(2-10) = 1.4\pm0.1\times10^{32}$\,ergs/s.  These
parameters are quite similar to our Abs PC TH model, with the
(significant) change that the 20\% moderate absorption seen in AM Her
is 100\% in the case of \ss17.  Of course, unlike the wide separation
in symbiotic systems, CVs are close binaries which show classical nova
outbursts \citep{Ibin96}.  However, the similarity of the hard X-ray
emission seen in \ss17 and AM Her suggests that the underlying
accretion processes are related in this object.

The \citet{Kennea07} study of symbiotic systems seen by the Swift BAT
(which includes \ss17) showed that these spectra usually showed X-ray
absorption that exceeded the optical extinction by orders of
magnitude, as was found here as well.  \citet{Kennea07} found a good
fit to the \swift XRT and BAT spectra using a fit similar to our
absorbed partial covering thermal model (``Abs PC TH''), with a
best-fit temperature similar to ours, although with a somewhat higher
partial covering fraction and somewhat lower local absorption.  The
\swift XRT data they used could only constrain a single iron line at
6.61 keV with width $0.18^{+0.21}_{-0.11}$\,keV, which they noted was
likely the sign of blended lines.  Our deeper observation shows this
is in fact the case, as \suzaku can easily resolve the three lines.

The three iron emission lines visible in Figures~\ref{fig:spec} and
\ref{fig:absspec} are all strong and narrow; none of our fits require
broadening beyond the instrumental resolution.  Although the \feHe and
\feH lines can be understood in the thermal models as due to the hot
plasma, the Fe K 6.4 keV line must be generated via fluorescence.
\citet{Ezuka99} showed that an 10 keV bremsstrahlung source embedded
in an absorber will generate (due to absorption and fluorescence into
the line of sight) a predicted equivalent width of the Fe K line of
0.67 (N$_{\rm H}/10^{24}{\rm cm}^{-2})$\,keV.  Using this result with
the ``Abs PC TH'' case (including both absorption components), we find
an equivalent width of 0.13 keV.  In addition, there may also be
fluorescence due to reflection from the surface of the compact object
(assumed here to be a white dwarf).  In this case, \citet{George91}
calculated that a 10-20 keV bremsstrahlung reflector subtending a
$2\pi$\ solid angle would generate an equivalent width of 0.1 keV,
assuming an observed angle to the reflecting surface of $60^{\circ}$.
We thus get a total equivalent width of 0.23 keV, consistent with the
actual value of $0.26^{+0.06}_{-0.04}$\,keV.  We also note, however,
that this calculation assumes that the reflected emission line is not
absorbed before escaping, so the calculated result is actually an
upper limit.  It seems likely therefore that the continuum around 6.4
keV is at least partially reflected, so our spectral models (which do
not include reflection) are incomplete.

With the existing data, we cannot distinguish (statistically) between
the power-law and thermal models after including a partial absorber.
Figure~\ref{fig:absspec} shows that the weak soft X-ray emission is
much better fit with the thermal model, although between 3-5 keV the
power-law model provides a better description.  Physically, the
thermal model represents an accretion column falling onto the surface
of the white dwarf.  This column is likely to have a range of
temperatures peaking around our single-temperature value
\citep{Ezuka99}.  Including a distribution of lower temperatures which
would improve the 3-5 keV fit, although more complex models are not
justified with the existing data.  In similar systems, such as the
symbiotic CH Cyg, the thermal origin of the plasma is confirmed by
multiple emission lines from ions such as Mg, Si, and S at lower
energies than the iron features.  The strong absorption seen here
obscures these lines; a longer or higher resolution observation is
needed to determine if in fact they are present.

The {\sl lack}\ of soft X-ray emission seen in \ss17, as compared to
CH Cyg, is also interesting.  Although similar to \ss17 in hard
X-rays, CH Cyg is $10-100\times$\ brighter than \ss17 in the soft
(0.5-2.0 keV) X-ray band \citep{Mukai07}.  \citet{Wheatley06} showed
that this soft X-ray emission from CH Cyg could be explained as
scattering in the ionized absorbing material surrounding the hard
X-ray source.  Their model is similar to that of a Seyfert 2 galaxy,
with a hard X-ray source surrounded by an edge-on torus of neutral
material with an ionized absorber above and below the torus.  This
anisotropic absorber removes any direct soft X-rays, but soft X-rays
emitted in other directions are still scattered into the line of sight
by a nearby absorber photoionized by the hard X-rays emitted by the
system.  The most significant difference between their hard X-ray
model and ours is that the partial covering fraction was $66\pm4$\%
rather than our value of $\ge85$\%.  Possibly in the case of \ss17 the
soft X-rays are simply absorbed by more distant neutral material, or
there is no ionized absorbing material above the accretion disk.  In
the former case, however, the relatively low Galactic absorption of
the system suggests this neutral material must still be quite near the
compact object.

Our initial observation of this source was done in the hope of
discovering a new ``highly-absorbed X-ray binary'' of the type
described by \citet{Kuulkers05}, and was driven by its BAT detection
and location near the Galactic plane.  Unlike the other symbiotic
systems that emit hard X-rays, \ss17 has apparently not shown any
signs of nova activity in the optical in the 100+ years since its
first description (although we emphasize that published observations
have been rather intermittent), and it has always shown weak soft
X-ray emission, particularly in our observation.  We cannot confirm
the nature of the companion, although it seems likely it is a white
dwarf based on the relatively low luminosity and iron lines similar to
other white dwarf symbiotics.  However, \ss17 may be the first example
of a system with significant Fe~K emission lines {\bf without}\
significant soft X-ray emission or optical outbursts.  This
characteristic is potentially important in explaining the Galactic
ridge X-ray emission (GRXE) which includes a hard continuum with broad
fluorescent Fe K, Fe~{\sc xxv}, and Fe~{\sc xxvi} iron lines in the
ratio 1:1.18:0.38 \citep{Koyama96}.  \citet{Mukai93} as well as
\citet{Revnivtsev06a,Revnivtsev06b} have suggested that the continuum
emission from the GRXE could be due to dwarf novae and
coronally-active stars, but did not explain the observed iron lines.
The ratio of the three iron lines strengths in \ss17 is 1:1:0.63
(errors of $\sim 15$\%).  If a large number of as-yet undiscovered
sources like \ss17 are common in the Galactic ridge, they could
explain the origin of its line emission as well.

\  \\ 
\noindent {\bf Acknowledgments }

We would like to acknowledge the Suzaku operations center at ISAS/JAXA
for their support as well as the HEASARC High Energy Archive at
NASA/Goddard Space Flight Center.  We are also grateful to Dr. Padi
Boyd and Dr. Lorella Angelini for helpful discussions.  This work was
supported by Suzaku Guest Observer Grant 344833.04.08.05.


\end{document}